\documentclass[useAMS,usenatbib,twocolumn]{mnras}
\usepackage{times}
\usepackage{url}
\usepackage[dvips]{graphicx}
\DeclareGraphicsExtensions{.pdf,.png,.jpg,.mps,.eps,.ps}
\usepackage{multicol}
\usepackage{subfig}
\usepackage{amsmath, array}
\usepackage{diagbox}
\usepackage{textcomp }

\usepackage{wrapfig}

\newcommand\flux{\ifmmode {\rm~erg cm}$^{-2}$\ ; {\rm s}$^{-1}$ \else ~erg cm$^{-2}$ s$^{-1}$\fi}
\newcommand\kms{\ifmmode {\rm~km\ s}$^{-1}$ \else ~km s$^{-1}$\fi}
\newcommand\Hunit{\ifmmode {\rm~km\ s}$^{-1}$\ {\rm Mpc}$^{-1}$
        \else ~km s$^{-1}$ Mpc$^{-1}$\fi}
\newcommand\ctssec{\ifmmode {\rm~count\ s}$^{-1}$ \else ~count s$^{-1}$\fi}
\newcommand\ergsec{\ifmmode {\rm~erg\ s}$^{-1}$ \else
        ~erg s$^{-1}$\fi}
\newcommand\funit{\ifmmode {\rm~erg\ s}$^{-1}$\ ; {\rm cm}$^{-2}$ \else
        ~ergs s$^{-1}$ cm$^{-2}$\fi}
\newcommand\phflux{\ifmmode {\rm~photon\ s}$^{-1}$\  ; {\rm cm}$^{-2}$
        \else   ~photon s$^{-1}$ cm$^{-2}$\fi}
\newcommand\efluxA{\ifmmode {\rm~erg\ s}$^{-1}$\ ; {\rm cm}$^{-2}$\ ; {\rm
        \AA}$^{-1}$ \else ~erg s$^{-1}$ cm$^{-2}$ \AA$^{-1}$\fi}
\newcommand\efluxHz{\ifmmode {\rm~erg\ s}$^{-1}$\ ; {\rm cm}$^{-2}$\ ; {\rm
        Hz}$^{-1}$ \else ~erg s$^{-1}$ cm$^{-2}$ Hz$^{-1}$\fi}
\newcommand\cc{\ifmmode {\rm~cm}$^{-3}$ \else cm$^{-3}$\fi}
\newcommand\fwhm{\ifmmode {\rm~FWHM} \else ${\rm~FWHM}$\fi}
\newcommand\msun{\ifmmode M_{\odot} \else $M_{\odot}$\fi}
\newcommand\lsun{\ifmmode L_{\odot} \else $L_{\odot}$\fi}

\newcommand\hbeta{\ifmmode {\rm H}\beta \else H$\beta$\fi}
\newcommand\kalpha{\ifmmode {\rm K}\alpha \else $K_\alpha$\fi}
\newcommand\lalpha{\ifmmode {\rm L}\alpha \else $L_\alpha$\fi}
\newcommand\nh{\ifmmode N_{\rm H} \else N$_{\rm H}$\fi}

\title[Blue straggler stars in NGC 2213]{Blue straggler stars beyond the Milky Way: a non-segregated population in the Large Magellanic Cloud cluster NGC 2213}
\author[C. Li \& J. Hong]
{Chengyuan Li$^1$\thanks{E-mail: chengyuan.li@mq.edu.au (CL); jongsuk.hong@pku.edu.cn (JH) }, Jongsuk Hong$^2$\\ 
$^1$Department of Physics and Astronomy, Macquarie University, Sydney, NSW 2109, Australia\\
$^2$Kavli Institute for Astronomy and Astrophysics, Peking University, Yi He Yuan Lu 5, HaiDian District, Beijing 100871, China\\
}

\date{Accepted XXX. Received YYY; in original form ZZZ}

\pubyear{2018}

\begin{document}
\label{firstpage}
\pagerange{\pageref{firstpage}--\pageref{lastpage}}
\maketitle

\begin{abstract}
Using the high-resolution observations obtained by the {\sl Hubble Space Telescope}, we analyzed the blue straggler stars (BSSs) in the Large Magellanic Cloud cluster NGC 2213. We found that the radial distribution of BSSs is consistent with that of the normal giant stars in NGC 2213, showing no evidence of mass segregation. However, an analytic calculation carried out for these BSSs shows that they are already dynamically old, because the estimated half-mass relaxation time for these BSSs is significantly shorter than the isochronal age of the cluster. 
We also performed direct $N$-body simulations for a NGC 2213-like cluster to understand the dynamical processes that lead to this none-segregated radial distribution of BSSs. Our numerical simulation shows that the presence of black hole subsystems inside the cluster centre can significantly affect the dynamical evolution of BSSs. The combined effects of the delayed segregation, binary disruption and exchange interactions of BSS progenitor binaries may result in this none-segregated radial distribution of BSSs in NGC 2213.
\end{abstract}


%====================================
\section{Introduction}
Blue straggler stars (BSSs) are massive main-sequence (MS) stars which should have evolved off in a star cluster. They were first discovered in the globular cluster (GC) M3 by \cite{Sand53a}, and are currently observed in all GCs properly studied \citep[e.g.,][]{Bald16a}. Two main leading scenarios are suggested being responsible for the BSSs formation: binary mass-transfer up to the coalescence of two components, and direct stellar collisions \citep{Mccr64a,Zinn76a,Hill76a}. 

It was claimed that GCs can be grouped into three distinct families in terms of the radial distribution of BSSs \citep{Ferr12a}: dynamically old GCs have their BSSs more concentrated than normal giant stars. As a result, the relative frequency between the BSSs and normal giant stars monotonically decreases from the cluster centre towards the periphery. Here {\it dynamically old} means the cluster is at least 5 times older than its half-mass relaxation time ($t_{\rm rh}$, see Supplementary Information of \cite{Ferr12a}). Less evolved GCs with dynamical age between one to five times the half-mass relaxation timescale have a bimodal distribution in the BSSs number fraction profile, which contains a peak in the cluster centre, a dip in the intermediate regions and another peak in the external regions \citep[e.g.,][]{Ferr97a,Li13a}. GCs that are still relaxing (younger than one half-mass relaxation time) show a completely flat relative frequency profile of BSSs \citep[e.g.,][]{Dale08a}, that is, their BSSs have a similar spatial distribution with normal stars. 

In addition, \cite{Ferr12a} found that for GCs with a bimodal radial distribution of BSSs, the position of the dip in the radial distribution corresponds well to the dynamical stage reached by their host clusters, thereby permitting a direct measure of the cluster dynamical age. This framework was subsequently cemented by recent observations \citep[e.g.,][]{Dale13a,Sann14a}. A recent numerical simulation run by \cite{2015ApJ...799...44M} provided additional support to the observations. However, it is also found that the bimodal radial distribution of BSSs is a very unstable feature \citep{Hypk17a}. \cite{2016ApJ...833L..29L,2016ApJ...833..252A} refined this dynamical clock by introducing a new parameter $A^{+}$, which represents the area enclosed between the cumulative radial distribution of BSSs and reference samples (usually evolved stars). They found that this parameter is a better indicator for measuring the dynamical evolution of globular clusters. 

% \cite{Ferr12a} suggested that the relationship between the morphology of BSSs radial distribution and the dynamical age of their host clusters can be explained by the dynamical friction and the consequent mass segregation process. Since BSSs are massive single stars or mass transferring binaries, they are more massive than bulk population stars on average. Dynamical friction drives these massive objects towards the cluster centre over time. Because in the inner region this dynamical timescale is much shorter than that in the outer region, BSSs in the central region would be segregated more rapidly than those in the cluster periphery. As a result, their radial distribution compared to normal stars can  develop a peak in the cluster centre and a dip in the intermediate regions that progressively moves outwards with time. This radial distribution will finally reach a monotonically decreasing shape once the entire cluster is fully relaxed. {\color{red} This framework that the shape of the radial distribution of BSSs can be used as a dynamical clock was subsequently cemented by the recent observations \citep[e.g.,][]{Dale13a,Sann14a} and numerical simulations \citep[e.g.,][]{2015ApJ...799...44M,2016ApJ...833L..29L}, however, was challenged by \cite{Hypk17a} recently. }

Recently, \citet{2016ApJ...833..252A} have found that the presence of a black hole (BH) subsystem plays a crucial role in altering the radial distribution of BSSs in star clusters. They suggested that the mass segregation of BSSs is significantly delayed by increasing the retention fraction of BHs inside clusters. The effects of the presence of BH subsystems on the early evolution of clusters already have been studied in literatures \citep[e.g.][]{1993Natur.364..423S,1993Natur.364..421K,2013MNRAS.432.2779B}. Due to significantly larger masses of BHs compared to other stellar components, BHs sink to the central region quickly and form a very dense system which is too dense not to be influenced by other stars outside. This subsystem undergoes the collapse in a very short time and generates large amount of energy to the entire cluster via the formation of binary and multiple BH systems. The energy generated from the BH subsystems also disturbs less massive stars like BSSs to be segregated to the central regions.

Although BSSs in old Galactic GCs (GGCs) are well explored, only few studies have been done for BSSs in young clusters \citep[e.g.,][]{Xin07a} or extragalactic GCs \citep[e.g.,][]{Li13a}. A key question of interest is whether our understanding of BSSs in old GGCs also can be applied to these stellar systems. The Large Magellanic Cloud (LMC) galaxy contains lots of clusters with a broad range of ages \citep[e.g.,][]{Mcla05a,Baum13a}. In addition, early studies \citep[e.g.,][]{1989ApJ...347L..69E,1991ApJS...76..185E,1992MNRAS.256..515E,2003MNRAS.338...85M} have found that there is a correlation between the core size and the age of young massive clusters, that the spread of core size increases with the cluster age. Later, by means of numerical simulations, \citet{2008MNRAS.386...65M} suggested that this increasing trend is due to the combined effects of stellar evolution mass-loss and the retention of BHs. If it is the case that some of young massive clusters in the LMC and the SMC still retain a significant number of BHs inside, the relaxation process of BSSs should be affected by the dynamics of the BH subsystem. 

In addition, in young star clusters, the dynamical disruption of binary systems  may be still ongoing \citep[e.g.,][]{Li13b,Gell13a}. It is also possible that the presence of a BH subsystem will further complicate the dynamical evolution of binaries and their products such as BSSs.

In this paper, we studied BSSs in the LMC cluster NGC 2213, which is an intermediate-age cluster with an isochronal age of $\sim$1.8 Gyr and a total mass of 36,000$M_{\odot}$ \citep{Baum13a}. We found that all the BSSs in NGC 2213 are dynamically old in terms of their relaxation time-scale. However, they do not significantly more segregated than normal giant stars, which is very different to BSSs in most dynamically evolved GGCs \citep[e.g.,][]{Ferr12a}. We also performed direct $N$-body simulations to understand this radial trend of BSSs obtained from the observation. Our numerical simulation shows that the combined effects of dynamical mass segregation and disruption of binary stars under the influence of a BH subsystem can produce a none-segregated population of BSSs. These processes are relatively less important for GGCs.

This article is organized as follows. Section \ref{S2} contains the details of observations and data reduction. Section \ref{S3} includes our main results. Section \ref{S4} is the physical discussions. We present the theoretical interpretations based on the numerical simulations in Section \ref{S5}. Finally, we summarize our study in Section \ref{S6}.
%====================================

%====================================
\section{Data reduction}\label{S2}
%====================================
\subsection{Photometry}
Our dataset was observed with the Hubble Space Telescope ({\sl HST}) Ultraviolet and Visible channel of the Wide Field Camera 3 (UVIS/WFC3). The resulting dataset is composed of a combination of six science images in the {\sl F475W} and {\sl F814W} filters with exposure times of 120 s + 600 s + 720 s and  30 s + 2$\times$700 s, respectively. Another dataset of a surrounding region  observed with the {\sl HST} Advanced Camera for Surveys (ACS) Wide Field Channel (WFC) was used as a reference field. The exposure times are 2$\times$500 s for each band. The program ID for both these two observations is GO-12257 (PI: L. Girardi). 

For UVIS/WFC3 images, we used the WFC3 module based on the {\sc dolphot2.0} package \citep{Dolp11a,Dolp13a} to perform the point-spread-function (PSF) photometry on the flat-field frames (named as `\_flt') of the scientific images. We first selected all detected objects flagged as `good stars' by the {\sc dolphot2.0} procedure. Then we removed all stars marked as `centrally saturated'. We finally adopted a filter employing the sharpness calculated by the {\sc dolphot2.0}, which enables us to filter out most sharp detections (e.g., cosmic rays) or extended sources (e.g., background galaxies). We only selected objects with $-0.3<$sharpness$<0.3$ in both frames. The {\sc dolphot2.0} package also provides a parameter called `crowding' for individual detected objects, which measures how much brighter the star would have been measured had nearby stars not been fit simultaneously (in units of mag). A high crowding value usually means that the object is poorly measured. For example, a detected object with a crowding of 0.1 mag means that it should be 0.1 mag fainter than the measurement. We also constrained all stars with their crowding no larger than 0.5 mag. For the ACS/WFC observations, we used the corresponding ACS module of the {\sc dolphot2.0} package to do the PSF photometry. The raw colour-magnitude diagram (CMD) of the NGC 2213 field is messy. After the same data reduction procedure as introduced above, lots of `junk' objects are removed and a much cleaner colour-magnitude diagram (CMD) can be obtained. However, because the BSSs are very bright. They all have very good photometry quality. We have confirmed that none of our selected BSSs will be removed followed by our data reduction process. For the observation of the reference field, we used the ACS module of the {\sc dolphot2.0} package to perform the same photometry and data reduction process. The CMDs derived from the UVIS/WFC3 images for the cluster field and the ACS/WFC images for the reference field, respectively, are presented in Fig.\ref{F1}.

\subsection{Isochrones fitting}
Using the PARSEC isochrones \citep{Bres12a}, we determined a best fitting isochronal age of log$(t/{\rm yr})$=9.26 ($\sim$1.8 Gyr) for the bulk stellar population of NGC 2213. The corresponding best fitting metallicity, distance modulus and extinction are $Z=0.006$\footnote{$Z_{\odot}=0.0152$}, $(m-M)_0$=18.50 mag and $E(B-V)$=0.06 mag, respectively. These results are close to the parameters derived by \cite{Baum13a} and \cite{grijs14a} (for the LMC distance modulus). Beyond the main-sequence turnoff (MSTO) region of the bulk stellar population, there are lots of stars which are significantly bluer and brighter than the MSTO stars. These stars are all BSS candidates. We found that these stars are all located in the region covered by two isochrones with ages of log$(t/{\rm yr})$=9.08 (1.2 Gyr) and log$(t/{\rm yr})$=8.48 (300 Myr) with the same metallicity, distance modulus and extinction.

\subsection{Determinations of stellar samples}
\begin{figure*}
  \centering
  \includegraphics[width=2.0\columnwidth]{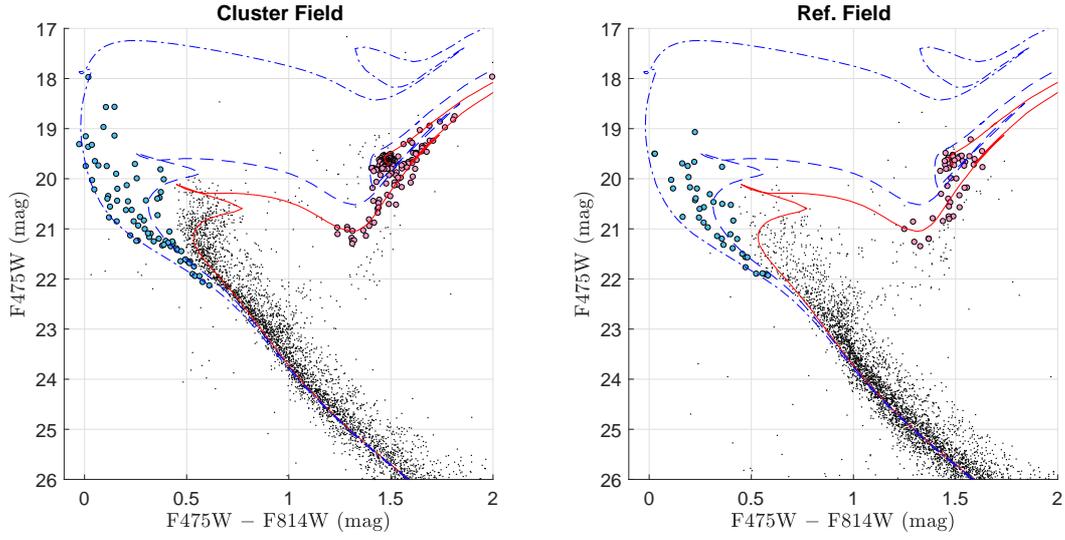}
  \caption{Left panel: The CMD of stars in the cluster region of NGC 2213 (UVIS/WFC3 image). Right panel: The CMD of stars in the reference field region (ACS/WFC image). The  selected BSSs (cyan circles) and giant stars (RGB, RC and AGB stars, pink circles) are highlighted. The red line is the isochrone with an age of log$(t/{\rm yr})$=9.26 (1.8 Gyr). The blue dashed line and blue dash-dotted line are isochrones with ages of log$(t/{\rm yr})$=9.08 (1.2 Gyr) and log$(t/{\rm yr})$=8.48 (300 Myr), respectively.}\label{F1}
\end{figure*}

We selected BSSs through the visual inspection by comparing their colour-magnitude distribution with the adopted two young isochrones (Fig.\ref{F1}). Stars located in the region covered by these isochrones or within the 3$\sigma$ region bracketing them (calculated by {\sc dolphot2.0}) are defined as BSSs (cyan circles in Fig.\ref{F1}). We also identified sample stars containing most red giant branch (RGB), red clump (RC), and asymptotic giant branch (AGB) stars. We first identified the position of the bottom of RGB stars in the CMD. Stars brighter than this position and located in the 3$\sigma$ (calculated by {\sc dolphot2.0}) region bracketing the isochrone are selected as sample stars. We confirmed that none of the detected BSS candidates and only a small fraction of RGB candidates ($\leq$10\%) have been removed after the data reduction. The selected BSSs and giant stars (RGB, RC \& AGB stars) are highlighted by cyan and pink circles in Fig.\ref{F1}, respectively.

From Fig.\ref{F1} we show that there are indeed lots of field stars which may contaminate the region where BSSs are located. We have detected 71 and 40 total BSS candidates in the cluster region and the reference field region, respectively. Because the size of the ACS/WFC field is about 1.55 times the UVIS/WFC3 field ($202''$$\times$$202''$ versus $162''$$\times$$162''$), that means about 26 stars of that 71 BSS candidates are likely field stars. The number of selected giant stars is 167 in the cluster region and 32 of them are likely field stars, statistically.

%%%%%%%%%%%%%%%%%%%%%%
% Fig. 1
%%%%%%%%%%%%%%%%%%%%%%

\subsection{Cluster global parameters}
\begin{figure*}
  \centering
  \includegraphics[width=2.0\columnwidth]{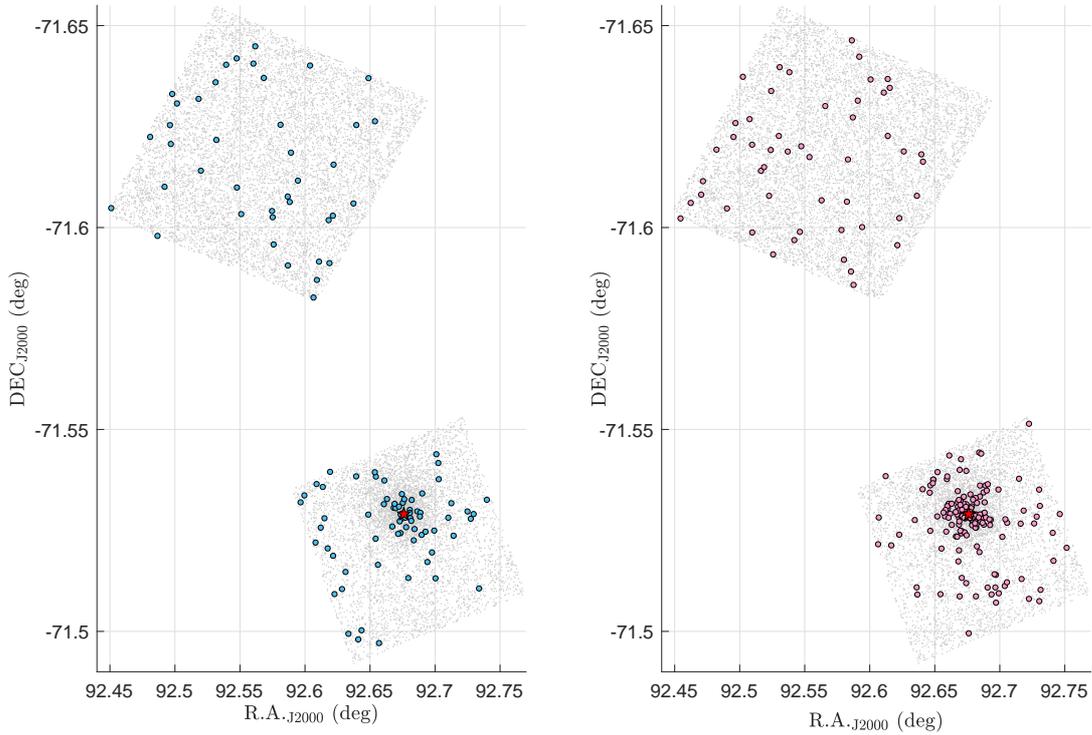}
  \caption{Left panel: The spatial distributions of all stars and BSSs (cyan circles). Right panel: The spatial distribution of all stars and giant stars (pink circles). The red pentagram indicates the cluster centre. The ACS/WFC footprint is located in the top, which represents the reference field. The WFC3/UVIS field is on the bottom, which contains the cluster.}\label{F2}
\end{figure*}

For all stars, we transformed their CCD coordinates (X,Y) to equatorial coordinates ($\alpha_{\rm J2000}$,$\delta_{\rm J2000}$) using the {\bf xy2rd} command of {\bf IRAF/STSDAS} task. 
We then investigated the stellar number density contour for the observed field. The position where the stellar number density reaches its maximum value is defined as the cluster centre. The resulting cluster centre coordinates are $\alpha_{\rm J2000}$=$06^{\rm h}10^{\rm m}42.24^{\rm s}$ and $\delta_{\rm J2000}$=$-71^{\circ}31'44.76''$, which are consistent with \cite{Baum13a}. In Fig.\ref{F2} we present the spatial distributions of all stars, selected BSSs as well as giant stars in both the fields of WFC3/UVIS (cluster region) and ACS/WFC (reference field). The spatial distribution of the selected BSSs shows an obvious sign of central concentration, which cannot be explained by a homogeneous distribution of field. 

%%%%%%%%%%%%%%%%%%%%%%
% Fig. 2
%%%%%%%%%%%%%%%%%%%%%%

\begin{figure*}
  \centering
  \includegraphics[width=2.0\columnwidth]{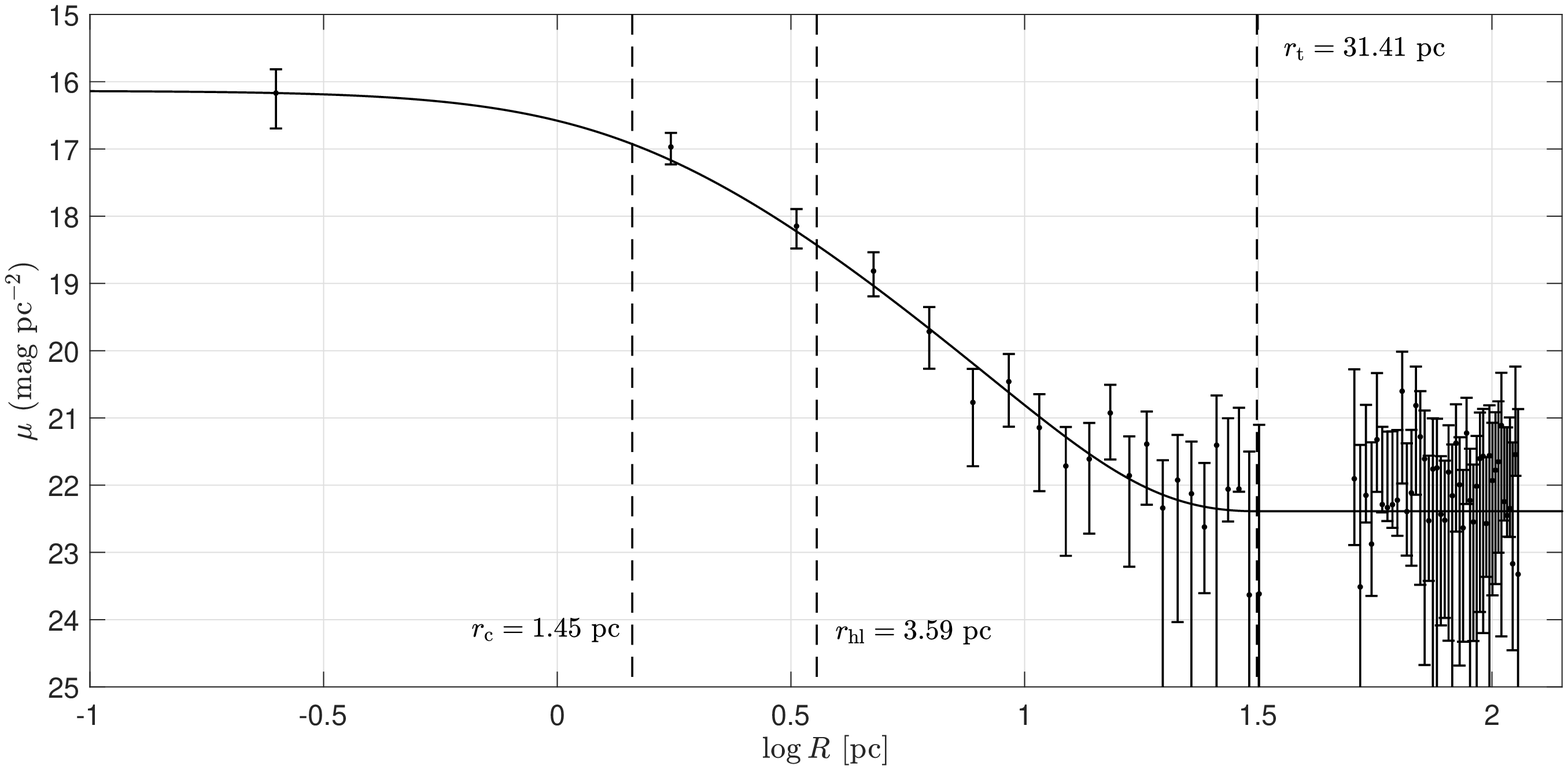}
  \caption{The brightness profile as well as the best fitting King model (black solid curve). Three dashed vertical lines from left to right indicate $r_{\rm c}$, $r_{\rm hl}$ and $r_{\rm t}$, respectively. Note there is a gap between $\log{R}$=1.5 to 1.7, which contains no data point. This is because we only have two separated observational fields (see Fig.\ref{F2})}\label{F3}
\end{figure*}

To obtain the brightness profile, we first selected all stars brighter than F475W=26 mag. 
Then, different annular rings from the cluster centre are determined with the intervals of 1 pc. In each ring, we calculated the total flux contributed from stars (in F475W band),  $f(r)=\sum{10^{({\rm F475W}-(m-M)_0)/(-2.5)}}$. The flux density is then $\rho(r)=f(r)/A(r)$, corresponding to a surface brightness of $\mu(r)=-2.5\log{\rho(r)}+(m-M)_0$ where $A(r)$ is the ring area calculated by Monte-Carlo method. We have also corrected the stellar flux incompleteness, which tells us how much lights are not detected in a given ring (see next subsection). We found that the flux completeness level for our selected stars ranges from 95\% to 100\%. This high flux completeness is expected because in a stellar system, bright stars contribute the majority of the flux and most of them are well detected. The resulting brightness profile is presented in Fig.\ref{F3}.

We used the King model with a constant representing the field brightness to fit the brightness profile \citep{King62a}
\begin{equation}
\mu'(r)=k\left[\frac{1}{\sqrt{1+(r/r_{\rm c})^2}}-\frac{1}{\sqrt{1+(r_{\rm t}/r_{\rm c})^2}}\right]+{b}
\end{equation}\label{E3}
where $r_{\rm c}$ and $r_{\rm t}$ are the core and tidal radii, $b$ is a constant which represents the contribution of the background, and $k$ is the normalization coefficient. Our best fitting core and tidal radii are $r_{\rm c}$=1.45$\pm0.02$ pc and $r_{\rm t}$=31.41$\pm3.02$ pc, respectively. The entire observed field of NGC 2213 is roughly equal to its tidal size\footnote{The radii for all detected stars in the observed cluster region do not exceed 34 pc.}. We therefore simply can take all stars in the whole UVIS/WFC3 image as cluster sample stars. The best fitting King profile is shown in Fig.\ref{F3} (the black solid line). Our derivation of $r_{\rm c}$ is consistent with \cite{Mcla05a} (1.52 pc). But our derived $r_{\rm t}$ is slightly smaller than that from \cite{Mcla05a} (41.68 pc). Since the tidal radius is only used to determine an appropriate reference field, we do not expect this difference as a major problem because the whole ACS/WFC image for the reference field is located well beyond 49.87 pc. Based on the derived brightness profile, we also calculated the half-light radius, which is $r_{\rm hl}=3.59^{+0.41}_{-0.59}$\footnote{The errorbar is simply the radial interval} pc. 
We summarize all global parameters in Table \ref{T1}.

%%%%%%%%%%%%%%%%%%%%%%
% Fig. 3
%%%%%%%%%%%%%%%%%%%%%%

\begin{table}
  \begin{center}
\caption{Derived global parameters for different stellar samples of NGC 2213.}\label{T1}
  \begin{tabular}{l l l}\hline
        &  All stars & BSSs\\\hline
   log ($t/{\rm yr}$) & 9.26 & 8.48--9.08 \\
$Z$ ($Z_{\odot}$=0.152) &0.006&0.006\\
$E(B-V)$ (mag) & 0.06 & 0.06 \\
$(m-M)_0$ (mag) & 18.50 & 18.50\\
$\alpha_{\rm J2000}$ & $06^{\rm h}10^{\rm m}42.24^{\rm s}$ & N/A\\
$\delta_{\rm J2000}$ & $-71^{\circ}31'44.76''$ & N/A\\
$r_{\rm c}$ (pc) & $1.45\pm0.02$ & N/A\\
$r_{\rm t}$ (pc) & $31.41\pm3.02$ & N/A\\
$r_{\rm hl}$ (pc) & $3.59^{+0.41}_{-0.59}$ & N/A\\\hline
  \end{tabular} 
  \end{center} 
\end{table} 

\subsection{Stellar completeness map}
Deriving reliable radial distributions for different stellar samples demands us to estimate the their completeness levels. In particular, different stellar samples that we are interested in include BSSs, giant stars, all stars and stars brighter than F475W=26 mag. For the completeness test, we generated a large number of artificial stars characterized by the same PSF as the real data. We added the artificial stars to the raw data image and made the detection of them using the same reduction procedure as for the real stars. We have generated two different artificial star samples defined by the 300 Myr- and 1.8 Gyr-old isochrones. For the young artificial stellar samples we only generated stars located in the region of BSSs. The total number of all artificial stars and young artificial stars are 2,625,600 and 272,000, respectively. We also generated a sample of artificial giant stars based on the same determination method that introduced above. The total number of artificial giant stars is 221,200. For all these artificial stellar samples, we divided them into two equal-size parts; a half of artificial stellar samples were added into the UVIS/WFC3 image and another half of them were added into the ACS/WFC image. All artificial stellar samples that we generated have a Kroupa-like mass function \citep{Krou01a}. We assumed that the spatial distributions of these artificial stars are homogeneous. To avoid a situation in which the artificial stars would increase the background and crowding level, we only added 100 artificial stars to the raw image each time and repeated this process 31,188 (26,256+2,720+2,212) times. 

To calculate the stellar completeness, we considered artificial stars that contribute to the sample incompleteness once they meet any of the criteria below,
\begin{enumerate}
\item They do not return any photometric result (with a magnitude of 99.99 in each passband).
\item They are not defined as `good star' by {\sc dolphot2.0}.
\item They are defined as `centrally saturated' by {\sc dolphot2.0}.
\item They have a crowding parameter larger than 0.5 mag.
\item Their sharpness parameter is not in between -0.3 and 0.3.
\end{enumerate}
Actually, this is exactly the data reduction process what we have performed to real stars. The stellar number completeness is defined as the number ratio between the stars that contribute the sample completeness and all input stars. Their light ratio is the stellar flux completeness, which has been used for correcting the stellar brightness profile (Fig.\ref{F3}).

\begin{figure*}
  \centering
  \includegraphics[width=2.0\columnwidth]{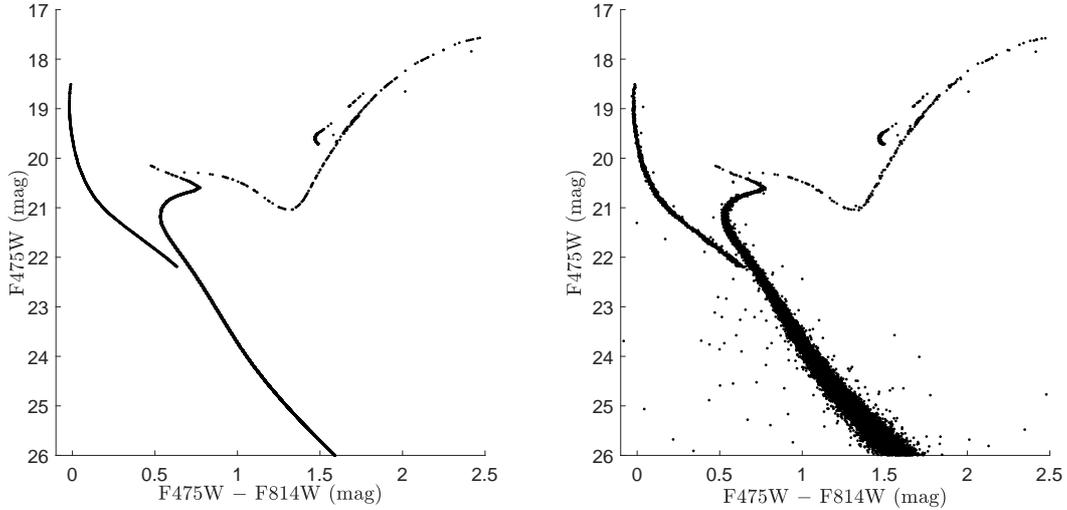}
  \caption{Left panel: The input CMD of artificial stars. The young MS branch represents the sample of artificial BSSs. Right panel: The output CMD of artificial stars based on our selection criteria. We only exhibit a small fraction of our artificial stars ($\sim$20,000) to avoid the overload of image space.}\label{F4}
\end{figure*}

\begin{figure*}
  \centering
  \includegraphics[width=1.8\columnwidth]{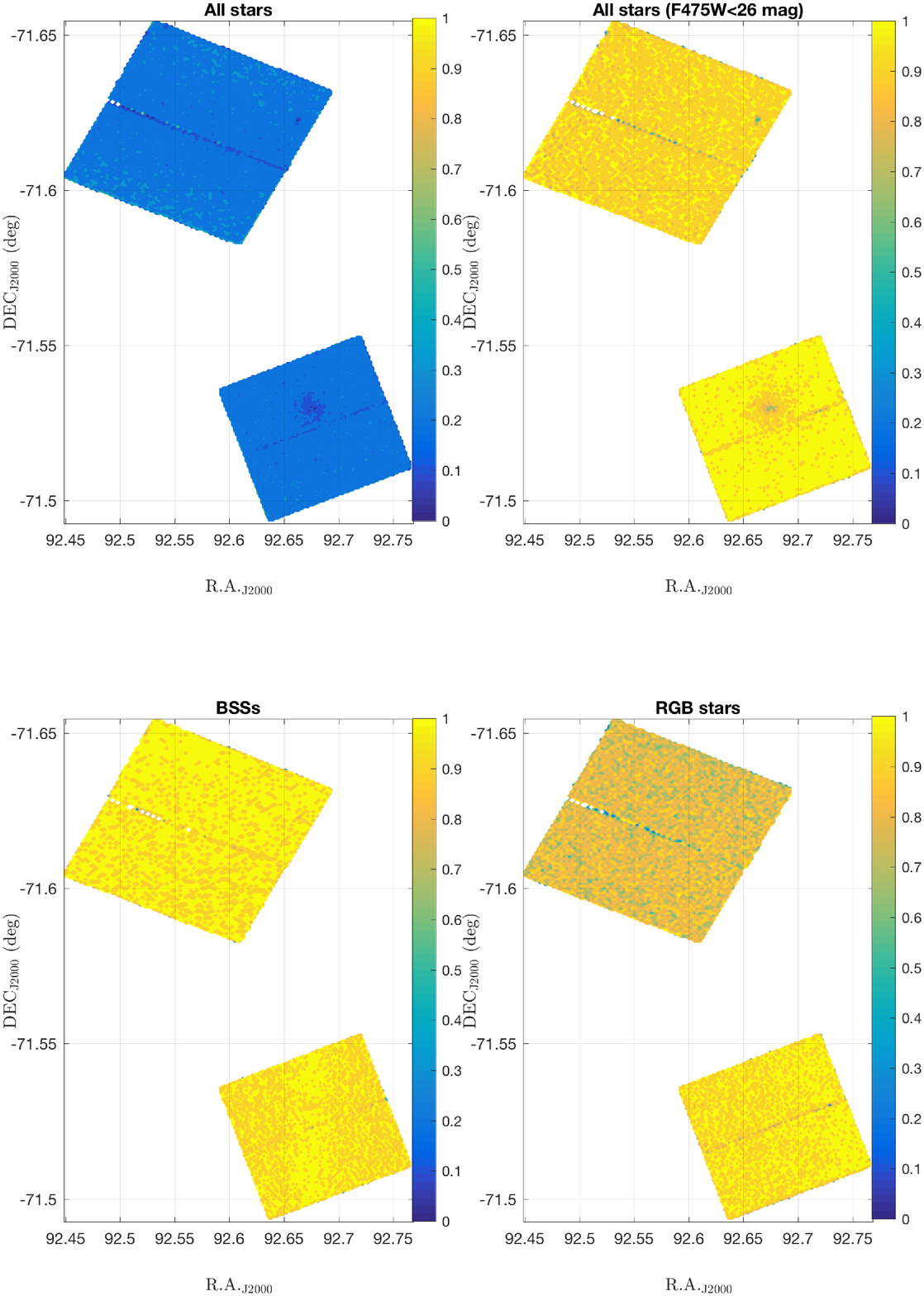}
  \caption{The stellar completeness map for all stars (top-left), all stars brighter than F475W=26 mag (top-right), BSSs (bottom-left) and giant stars (bottom-right)}\label{F5}
\end{figure*}

In Fig.\ref{F4}, we show examples of both the input and output CMDs for our artificial stellar samples. In Fig.\ref{F5}, we show the stellar number completeness maps for (1) all stars; (2) all stars brighter than F475W=26 mag; (3) BSSs and (4) giant stars. 

%%%%%%%%%%%%%%%%%%%%%%
% Fig. 4
%%%%%%%%%%%%%%%%%%%%%%

%%%%%%%%%%%%%%%%%%%%%%
% Fig. 5
%%%%%%%%%%%%%%%%%%%%%%

As shown in Fig.\ref{F5}, except for the stellar completeness map for all stars. The completeness level for other stellar samples are all higher than at least 80\%. The low level of completeness for all stars is expected, because our artificial stellar sample contains a large number of low-mass stars based on the adopted Kroupa mass function. Almost all of these low-mass stars are unable to be detected, therefore contributing the incompleteness (the corresponding magnitudes for the lowest mass of the artificial stars are F475W=33.35 mag and F814W=29.98 mag). The average completeness levels for the cluster region and the reference field are different. This is because the observed images for the cluster and the reference field have different exposure times. The difference in their completeness will be corrected in our follow up analysis.

\section{Results}\label{S3}
\begin{figure}
  \centering
  \includegraphics[width=1.0\columnwidth]{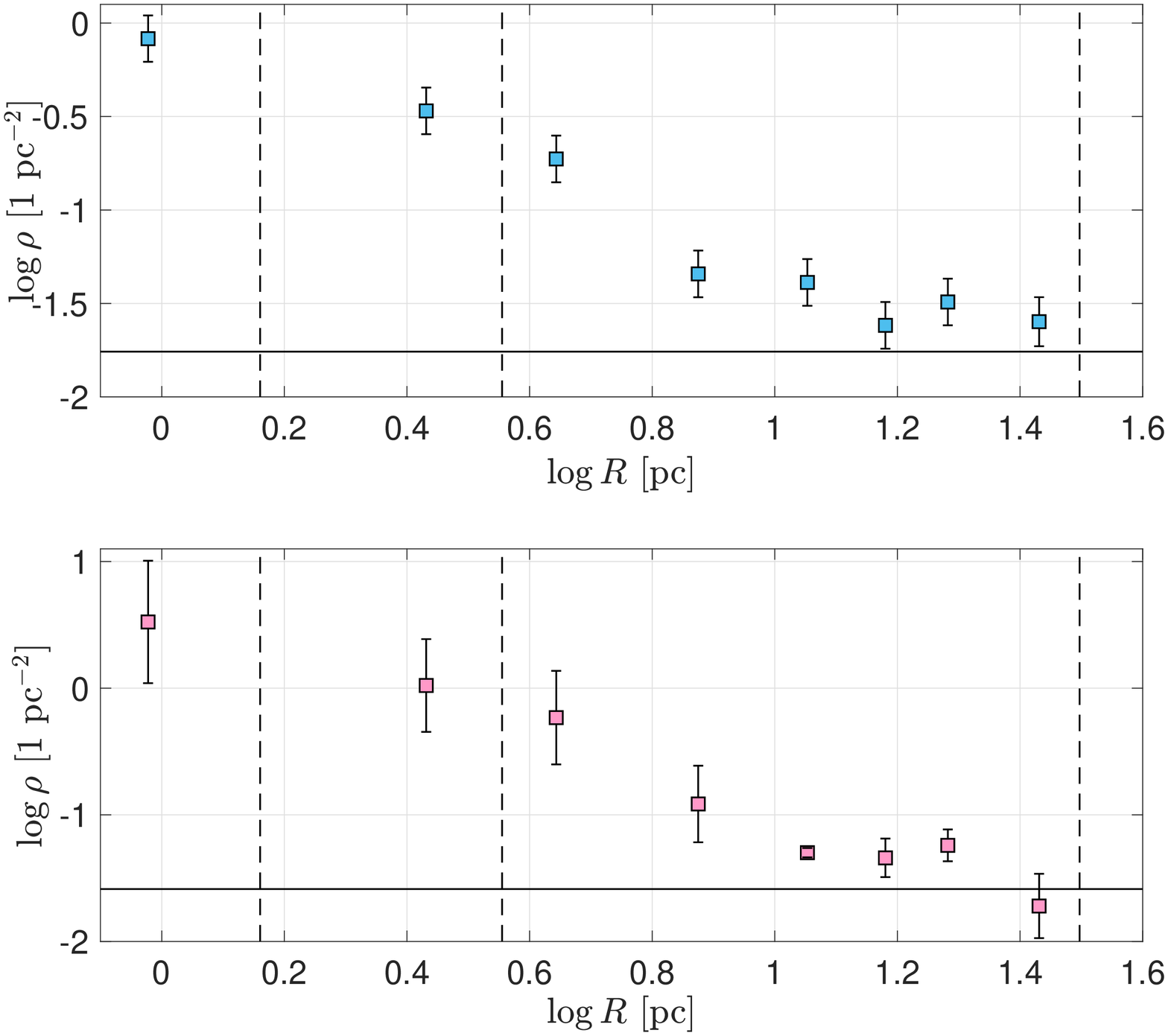}
  \caption{Number density profiles of BSSs (top) and giant stars (bottom). Three vertical dashed lines represent $r_{\rm c}$, $r_{\rm hl}$ and $r_{\rm t}$, respectively. The bottom horizontal lines represent the field densities of BSSs and giant stars.}\label{F6}
\end{figure}

In Fig.\ref{F6} we present the number density profiles of BSSs and giant stars, with stellar completeness corrected. i.e., 
\begin{equation}
\rho(r)=\frac{N(r)}{A(r)c(r)}
\end{equation}
where $N(r)$ and $c(r)$ are number of stars and the corresponding number completeness in the ring centred at radius $r$, $A(r)$ is the ring area. Because both the BSSs and giant stars are bright, their average completeness levels are very high (see Fig.\ref{F5}). The completeness variation at different radius would not dramatically change the morphology of the number density profiles for the BSSs and giant stars. The number density profiles for both the BSSs and giant stars exhibit an apparent peak in the cluster central region, which again support that these stellar samples are dominated by cluster members rather than a sample of unique field stars. Their number densities within the tidal radius are systematically higher than the field level. 

%%%%%%%%%%%%%%%%%%%%%%
% Fig. 6
%%%%%%%%%%%%%%%%%%%%%%

We then explore the number ratio between the BSSs and giant stars along with radius,
\begin{equation}
f_{\rm B/G}(r)=\frac{N_{\rm B}(r)}{N_{\rm G}(r)}=\frac{N_{\rm B}^{\rm obs}(r)/c_{\rm B}(r)-\alpha(r)N_{\rm B,f}^{\rm obs}/c_{\rm B,f}}{N_{\rm G}^{\rm obs}(r)/c_{\rm G}(r)-\alpha(r)N_{\rm G,f}^{\rm obs}/c_{\rm G,f}}
\end{equation}
Here $N_{\rm B}^{\rm obs}(r)$, $N_{\rm G}^{\rm obs}(r)$ are numbers of observed BSSs and giant stars at radius $r$. $c_{\rm B}(r)$ and $c_{\rm G}(r)$ are the corresponding stellar number completeness. $N_{\rm B,f}$ and $N_{\rm G,f}$ are numbers of observed BSSs and giant stars in the reference field region. $c_{\rm B,f}$ and $c_{\rm G,f}$ are the corresponding stellar number completeness of the reference field. $\alpha(r)$=$A(r)/A_{\rm ACS}$ is the area ratio between the ring and the reference field. 

\begin{figure}
  \centering
  \includegraphics[width=1.0\columnwidth]{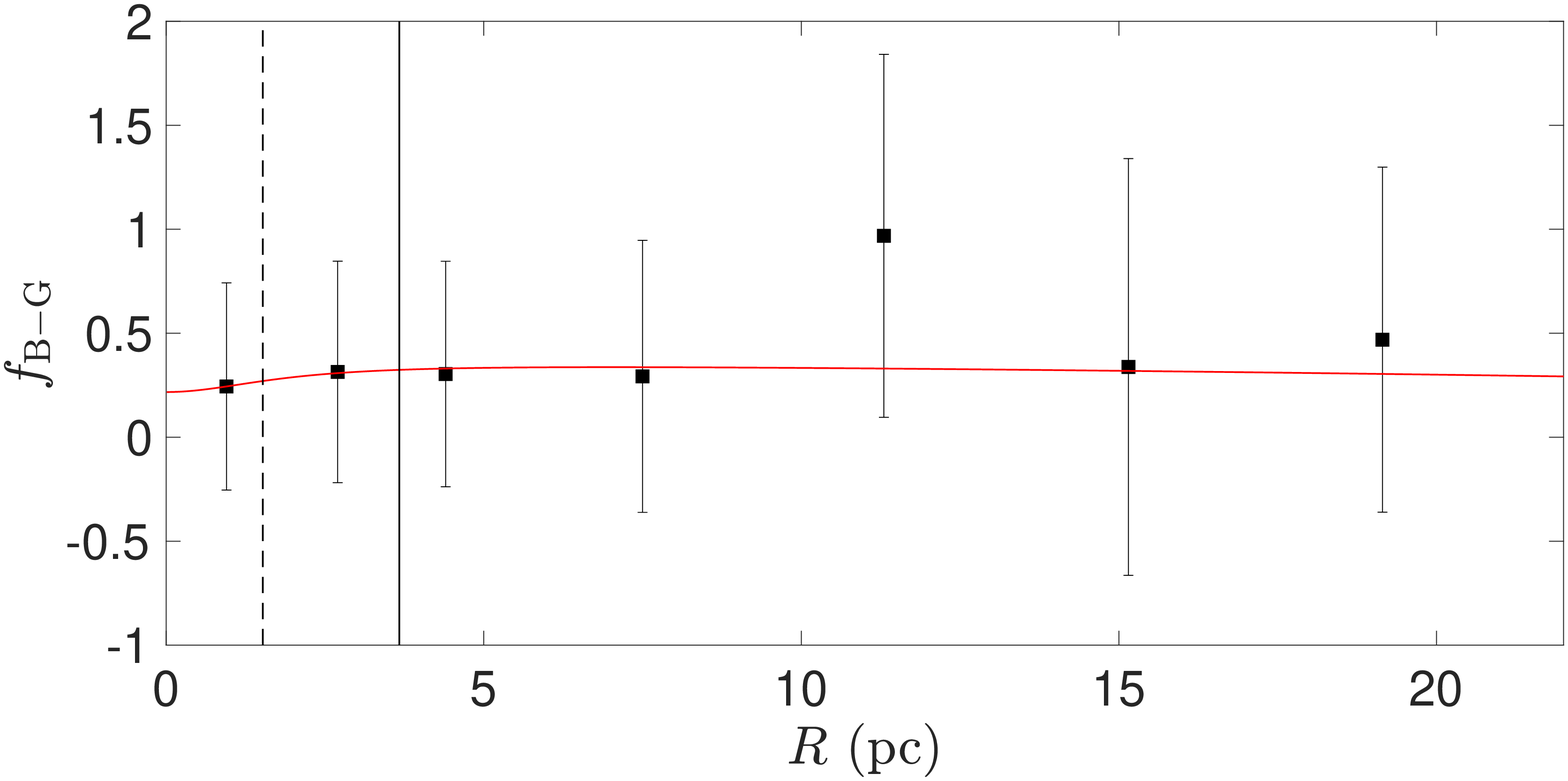}
  \caption{The profile of number ratio between BSSs and giant stars. The red solid line represents the ratio between the best fitting King models for BSSs and giant stars. The vertical dashed line and solid line are $r_{\rm c}$ and $r_{\rm hl}$, respectively.}\label{F7}
\end{figure}

Fig.\ref{F7} shows that the radial distributions for BSSs and giant stars are not significantly different. This is reflected by the profile of their number fraction, which remains unchanged from the cluster centre to the outer region. This profile can be described by the ratio between their best fitting King models, i.e.,
\begin{equation}
f(r)=\frac{K\left[\frac{1}{\sqrt{1+(r/R_{\rm c})}}-\frac{1}{\sqrt{1+(R_{\rm t}/R_{\rm c})}}\right]}{K'\left[\frac{1}{\sqrt{1+(r/R'_{\rm c})}}-\frac{1}{\sqrt{1+(R'_{\rm t}/R'_{\rm c})}}\right]}
\end{equation} 
For clarity, we do not directly use this formula to fit the number fraction profile in Fig.\ref{F7} (otherwise we will use a six parameter model to fit only seven data points.). Instead, we use the King model to fit independently the number density profiles of BSSs and giant stars (as shown in Fig.\ref{F6}) and then calculate their ratio. The ratio between their best fitting King models also indicates that there is no significant difference in the radial distribution of BSSs and giant stars (the red solid line in Fig.\ref{F7}). 

%%%%%%%%%%%%%%%%%%%%%%
% Fig. 6
%%%%%%%%%%%%%%%%%%%%%%

To examine whether BSSs are dynamically evolved, we evaluated the half-mass relaxation time for BSSs through the formula introduced by \cite{Meyl87a}
\begin{equation}
t_{\rm rh}=8.92\times10^5\frac{M_{\rm tot}^{1/2}}{{m}}\frac{r^{3/2}_{\rm hm}}{\log{(0.4M_{\rm tot}/{m})}}\,yr
\end{equation}
where $M_{\rm tot}$ is the total mass of the cluster, which is $\sim$36,000 $M_{\odot}$ \citep[see][]{Baum13a}, ${m}$ and $r_{\rm hm}$ are the typical mass for stars of interest in units of $M_{\odot}$ and the half-mass radius in pc, respectively. Because BSSs do not belong to the bulk population, it is very difficult to constrain their masses simply through the photometry. We estimated the mass range of these BSSs through the adopted two young isochrones, which is $\sim$1.35$M_{\odot}$--3.05$M_{\odot}$. This corresponds to $\sim$0.85--1.91 times the mass of MSTO stars. We simply assumed that the half-mass radius is equal to the half-light radius for this calculation, i.e., $r_{\rm hm}=r_{\rm hl}=$3.59 pc. The resulting half-mass relaxation time for BSSs is $\sim$100--210 Myr. If these BSSs are products of binary evolution or stellar collisions of the main population stars of NGC 2213 with the isochronal age of $\sim$1.8 Gyr, their dynamical age should be $\sim$9--18$t_{\rm rh}$. Therefore, BSSs in NGC 2213 are expected to be dynamically relaxed.

%====================================

%====================================
\section{Discussions}\label{S4}
%====================================
We first examined whether stellar collisions can efficiently produce the observed BSSs. We use the formula introduced by \cite{Davi04a} to evaluate the expected number of collisional BSSs formed within last 1 Gyr,
\begin{equation}
N_{\rm col}=0.03225\frac{f^2_{\rm mms}N_{\rm c}n_{\rm c,5}r_{\rm col}m_{\rm BSS}}{V_{\rm rel}}
\end{equation}
where $f_{\rm mms}$ is the fraction of MS stars in the core which are massive enough to form a BSS after the collision, $f_{\rm mms}=0.25$ is previously suggested \citep[see e.g.,][]{1995MNRAS.276..876D}. $N_{\rm c}$ is the number of stars contained in the cluster core region. To obtain the number of stars in the core region, we calculated the number of stars with masses between $\sim$1.0$M_{\odot}$ and $\sim$1.6$M_{\odot}$. The expected number of stars within this mass range in the core is $\sim$145 (with stellar number completeness corrected). We then assumed that these stars follow a Kroupa mass-function \citep{Krou01a} and then evaluated the total number of stars by extrapolating this mass-function down to 0.08$M_{\odot}$. The expected number of stars located in the cluster core is roughly 3,400. $n_{c,5}$ is the core density of stars in units of $10^5$ pc$^{-3}$, which is the number ratio between stars contained in the core and the core volume, yielding $n_{c,5}\sim0.003$. $r_{\rm col}$ is the minimum separation of two colliding stars in units of the solar radius, $R_{\odot}$. Assuming the average mass of BSSs is $m_{\rm BSS}\sim2.2$ $M_{\odot}$. A MS star with this mass would have a radius of $r\sim$2.1 $R_{\odot}$ \citep{Demi91a}. We simply adopted 4 $R_{\odot}$ for $r_{\rm col}$ (about twice the stellar radius). $V_{\rm rel}$ is the relative incoming velocity of binaries at infinity, which can be written as, 
\begin{equation}
V_{\rm rel}=\sqrt{2}\sigma=\sqrt{\frac{4GM_{\rm c}}{r_{\rm c}}}
\end{equation}
where $\sigma$ is the stellar velocity dispersion in the core region. $M_{\rm c}$ is the stellar mass of the cluster core, which is $\sim1,200$ $M_{\odot}$ based on our calculation. The resulting central stellar velocity dispersion for NGC 2213 is $\sigma\sim$2.7 km s$^{-1}$. Finally we obtained that only $\sim$0.08 collisional BSSs could form within past 1 Gyr. Note that this number only represents the number of BSSs formed through collisions between two single stars. Some high-order stellar interactions, like binary--single star and binary--binary interactions, may increase the production rate of collisional BSSs \citep[see e.g.,][]{2013MNRAS.428..897L}. As suggested by \cite{Davi04a}, the realistic value of collisional rate involving high-order stellar interactions should be still of the same order. Since there are at least 45 (71$-$26) BSSs have been detected in the cluster region of NGC 2213. The only viable interpretation of the origin of these BSSs is therefore binary evolution. 

However, if these BSSs are directly formed through the evolution of binary stars, they should be all dynamically old with their dynamical ages at least 10 times greater that their half-mass relaxation timescale as we evaluated in the previous section. According to the Supplementary Information of \citet{Ferr12a}, a dynamically old BSS population should exhibit an obvious central peak in their number fraction profile unlike BSSs in NGC 2213. This discrepancy might be due to different dynamical processes between young and old clusters. 

If the observed BSSs in NGC 2213 are not rejuvenated stars by binary evolution of existing populations stars but actually young MS stars formed more recently, 
it is possible that they were initially less segregated than normal giant stars. 
\citet{Bekk17a} suggested that during the hierarchical star cluster formation within fractal giant molecular clouds, younger stars can from using the AGB ejecta from unbound star clusters and be subsequently accreted to the most massive star cluster which is the progenitor of a GC.
In this case, young generation stars can be less concentrated that older generation stars.
In addition, minor merger(s) of star clusters also can result in the less concentrated radial distribution of younger generation stars \citep[e.g.,][]{Hong17a}. 
However, both hierarchical formation and merger scenarios cannot explain the large age variation of BSSs in NGC 2213. The merger scenario \citep[e.g.,][]{Hong17a} needs multiple merger events to reproduce this age distribution, which is highly unlikely, and the hierarchical formation scenario only reproduces the age difference less than $\sim$100 Myr \citep[see also the section 2.6 in][]{Bekk17a}. It is also puzzling why the maximum stellar mass for these young stars does not exceed two times the mass of turnoff stars. Indeed, the mass limit of these BSSs is a strong evidence which support that they are binary products. We therefore do not see these scenarios as potential explanations on the origin of these non-segregated BSSs.

A promising explanation is that the dynamical disruption of binary systems in NGC 2213 is still ongoing  while the disruption of BSS progenitor binaries is not very important in old clusters \citep[see e.g.,][]{2013AJ....145....8G}. \cite{Gell13a} have calculated the binary segregation in young massive clusters with the initial binary disruption being considered. They found that within the first two half-mass relaxation time of a cluster, binaries would not develop a obvious central peak in their frequency profile. Since all of our detected BSSs are likely binary products, it is therefore not strange that the observed BSSs does not show a significant sign of mass segregation if their progenitor binary systems in the inner regions are preferentially disrupted. In addition, the presence of a BH system may have strong impact on the radial distributions of BSSs \citep{2016ApJ...833..252A,2016MNRAS.462.2333P}. BH subsystems would segregate to the cluster centre much more rapidly than other stellar populations, disturbing these less massive populations of stars, including BSSs. The numerical modeling illustrating the effects of internal dynamics on shaping the frequency profiles of BSSs is following in the next section.

\section{Theoretical interpretations\label{S5}}

To examine this preliminary idea, we design a numerical simulation to study the radial behaviour of BSSs in a NGC 2213-like stellar system. We used the {\sc nbody6++gpu} \citep{2015MNRAS.450.4070W}, which is one of the most recent versions of {\sc nbody} \citep{2003gnbs.book.....A} direct $N$-body code taking the advantages of recent developments of hardware including GPU acceleration.
 We considered the initial mass function based on \citet{Krou01a} with the mass range from 0.08 to 100 $M_{\odot}$. These stars with different masses evolve inside the code based on the stellar evolution recipes \citep{2000MNRAS.315..543H,2002MNRAS.329..897H}. The mass fallback mechanism \citep{2002ApJ...572..407B} for BHs which leads to the retention of significant fraction of BHs was applied. To track the dynamical evolution of BSS candidates, we also took some fraction of primordial binaries into account.

The initial condition of the simulation was chosen to reproduce the observational surface density profile of NGC 2213 at the age of 1.8 Gyr. We used \citet{1966AJ.....71...64K} model with the initial concentration parameter $W_0 = 7$.
The initial number of stars in our cluster model is 120,000 including 24,000 primordial binaries. The initial orbital properties of primordial binaries follow the period and eccentricity distribution based on \citet{1995MNRAS.277.1507K}. The initial total mass is 6.90 $\times10^{4} M_{\odot}$ and the initial half-mass radius is $\sim$2.4 pc (in 3-dimensional; $\sim$1.8 pc in projection). We also considered the strongest tidal effect from nearby galaxies (the Milky Way, the LMC and the SMC) and the initial tidal radius is 42 pc. After 1.8 Gyr calculations, the cluster mass is about 3.38 $\times10^{4} M_{\odot}$. At this time, the 3-dimensional half mass radius is $\sim$6.5 pc including all stellar components and the tidal radius is $\sim$31.8 pc. The total number of BHs formed in this simulation is 232 and 120 of them are immediately ejected from the cluster due to the natal kick induced by the supernova explosion. The mass of the most massive BH is $\sim$18$M_{\odot}$ \citep[see e.g.,][for more details; note that the mass distribution and the retention fraction of BHs are sensitive to the metallicity]{2016MNRAS.463.2109R}.  At 1.8 Gyr, the number of remaining BHs is 56 as BHs escape from the cluster by the dynamical interactions\footnote{There are 8 escaping BH binaries formed dynamically in the cluster which can be potential gravitational wave sources \citep[see e.g.,][]{2017MNRAS.467..524B,2018MNRAS.473..909B}.}

\begin{figure}
  \centering
  \includegraphics[width=1.0\columnwidth]{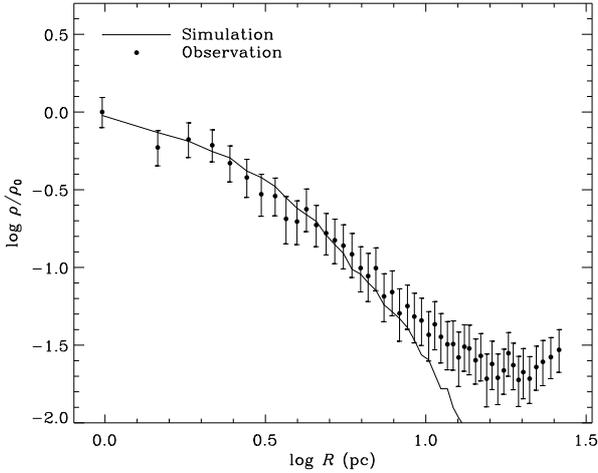}
  \caption{{\it Top panel}: Surface number density profiles normalized to the central values for the simulation model and the observational data. The completeness is corrected for the number density profile from the observational data. The error bars for the surface number density profile from the observation represent the Poissonian error, i.e., $\delta{\rho}=\sqrt{N}/A$}\label{F8}
\end{figure}

Fig. \ref{F8} shows the radial profiles of the surface number density from the simulation and the observation at the current time. To be consistent with the observations, we obtained the surface number density profile for the simulation results by only using visible stars more massive than $m > 0.6M_{\odot}$, where the lowest mass is from the observational cut-off (F475W = 26 mag).  The surface number density profile from the simulation agrees overall with that obtained from the observation. The discrepancy at the large radii ($R > 10$ pc) may be due to the incomplete subtraction of field stars of observational data (see also Fig. \ref{F3}; the brightness density at the outer regions is comparable to that of stars in the reference frame outside the tidal radius.). The core radius defined as the radius where the surface number density becomes a half of the central surface number density is $\sim$2.5 pc. This value is larger than that obtained from the best fitting of the observed surface brightness profile as shown in Fig. \ref{F3}. This is because the mass segregation process induces the inconsistency between the number and luminosity density profiles. 

In this numerical study, we do not consider the actual formation of BSSs in the simulations because we need much larger fraction of primordial binary to obtain the sufficient number of BSSs for better statistics, which is very difficult to be dealt with by the {\sc nbody6++gpu} code currently. Instead, we follow the dynamical evolution of binary stars that can be progenitors of BSSs during binary stellar evolution. We assumed that the collisional formation of BSSs in NGC 2213 is negligible as we calculated the BSSs' formation rates by stellar collisions in the previous section. In addition, \citet{2013ApJ...777..106C} found that there is a very weak correlation between the cluster encounter rates and the number of BSSs in GGCs. Therefore the radial distribution of BSSs is more affected by the dynamical evolution of binaries than the radial variation of stellar encounter rates.

\begin{figure}
  \centering
  \includegraphics[width=1.0\columnwidth]{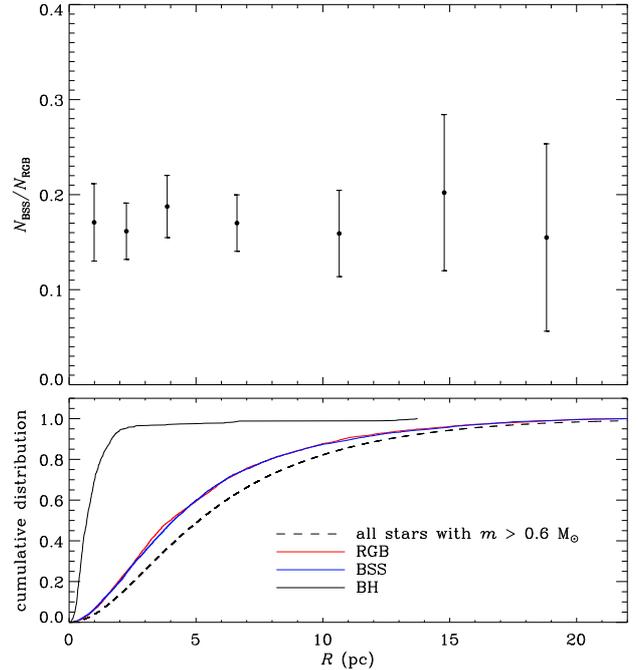}
  \caption{{\it Top panel}: Radial profile of the number ratio of BSSs to RGB stars from the simulation model. The error bars represent the combined Poissonian error of the numbers of BSSs and RGB stars. {\it Bottom panel}: Cumulative number distribution of stars with $m > 0.6 M_{\odot}$ that corresponds to the observational limit, RGB stars, BSSs and BHs, respectively. Five snapshots around $T = 1.8$ Gyr have been combined for the profiles with better statistics.}\label{F9}
\end{figure}

Top panel of Fig. \ref{F9} shows the radial profile of the frequency of BSSs defined as the ratio of the number of BSSs to the number of RGB stars in the projected radii bins. As we mentioned above, we do not count the number of actual BSSs but count the number of binaries with the sum of the masses of the progenitor MS stars in binary systems in between 1.7 $M_{\odot}$ and 3.2 $M_{\odot}$, where the MSTO mass is assumed to be $\sim$1.6 $M_{\odot}$. 
Although we estimated the mass of BSSs roughly down to 1.35 $M_{\odot}$ based on the isochrone fitting, here we exclude binaries with the mass below 1.7 $M_{\odot}$ because the number fraction of BSSs among binaries in that mass range is very small. In order to match the frequency of BSSs from the simulation model with that from the observation, we adopted 10 per cent of an artificial probability of BSSs among binary stars in the mass range. Note that this artificial treatment should be done in caution and needed to be justified by binary stellar evolution models for the formation of BSSs. We simply assumed that this probability is independent on the radial distance from the centre due to the inefficient formation of BSSs by stellar collisions.
Even if the dynamics does matter, the weak perturbations that shake the orbital properties of binaries and eventually lead to the formation of BSSs through the binary mass transfer or binary merger can take place in low density regions without significant differences in the rates \citep[see also the Fig. 3 in ][]{2017MNRAS.464.2511H}. In the top panel of Fig. \ref{F9}, the radial profile of the number ratio of BSSs to RGB stars infers that there is no significant discrepancy in the radial distribution of BSSs compared to that of normal RGB stars as similar to that shown in the observational data (Fig. \ref{F3}).

To better illustrate, we show the cumulative number distributions of different groups of stars in the bottom panel of Fig. \ref{F9}. The cumulative number distribution of BSSs shows an evidence of mass segregation of BSSs by comparison with the distribution of all observable stars while
the cumulative distribution of BSSs is almost identical with that of RGB stars as shown in the radial profile of BSS frequency.
On the other hand, the cumulative distribution of BHs shows that BHs are extremely concentrated to the central regions: the half number radius of BHs is $\sim$5 times smaller than that of BSSs or RGB stars and 90 per cent of BHs reside within 1.5 pc.

There could be some dynamical processes responsible for the none-segregated radial distribution of BSSs in NGC 2213-like clusters. From the equipartition theory, high-mass stars lose the energy and angular momentum by the dynamical interactions with low-mass stars and fall to the central regions \citep{2008gady.book.....B}. However, if there is a dense BH subsystem located in the innermost region like our simulated cluster, stars in the inner regions are more likely to interact with BHs due to larger cross sections by the gravitational focusing.
Since the mass difference between BSSs and RGB stars is relatively small compared the the difference with masses of BHs, the dynamical evolution of BSSs and RGB stars such as the mass segregation might not be significantly different \citep[see also][]{2016ApJ...833..252A}. 
It is very obvious that BSS progenitor binaries with small binding energy (i.e., soft binaries) are easily disrupted by the interactions with other stars and binaries \citep{2003gmbp.book.....H}. However, even hard BSS progenitor binaries can be depleted in the central regions with a BH subsystem. Classical studies for binary interactions \citep[e.g.,][]{1996ApJ...467..359H} suggested that one of the binary components is more likely to be replaced by an encountering star during binary-single interactions when the encountering star is much more massive than the masses of original binary components. If a BSS progenitor binary segregated to the centre encounters a BH, this binary might become a BH-MS binary as soon as encountered with a BH. These combined effects of delayed segregation, binary disruption and binary exchange events can produce a none-segregated distribution of BSSs.

%====================================
\section{Summary and Conclusions}\label{S6}	
Based on the high-precision multi-band HST photometry, we analyzed the BSSs in the LMC cluster NGC 2213. Our main results and conclusions are summarized below:
\begin{itemize}
\item[*] We totally detected 71 stars located in the region of BSSs of the cluster. We evaluated that about 26 of them are probably field stars. Therefore most of them are indeed cluster members. In addition, the overall spatial distribution of BSSs also shows an obvious pattern of central concentration, indicating that a majority of them are genuine cluster members. 
\item[*] Our analytic calculation shows that the BSS formation by stellar collisions does not account for observed number of BSSs. We suggested that the binary evolution channel is responsible for the presence of these BSSs.
\item[*] The half-mass relaxation times for these BSSs are very short, which does not exceed $\sim$210 Myr. These BSSs are expected to be dynamically relaxed if they are the products of binary evolution. However, these BSSs are not more segregated than normal giant stars. The number fraction profile between them does not significantly change with radius.
\item[*] We performed direct $N$-body simulations for NGC 2213-like clusters to understand the origin of the none-segregated distribution of BSSs. We investigated the dynamical evolution of binaries that can be the progenitors of BSSs and found that the presence of a BH subsystem inside star clusters can significantly affect the radial distribution of BSSs in several ways (delaying the segregation of BSSs, destroying BSS progenitor binaries and replacing BBS progenitor binaries' components). 
\item[*] Li et al. (in preparation) analyzed 11 young and intermediate age clusters in the LMC and found that individual clusters have different radial distributions of the number fraction of BSSs without any strong correlation with cluster physical properties (e.g., mass, core, half-mass radii, etc.). More comprehensive numerical simulations are therefore needed for better understanding if the observed radial distribution of BSSs is indeed related to the dynamical behaviors of the BH subsystem.
\end{itemize}
%====================================

\section*{Acknowledgements}
We thank the anonymous referee for his/her valuable comments. CL acknowledges funding support from the Macquarie Research Fellowship Scheme. JH acknowledges support from the China Postdoctoral Science Foundation, Grant No. 2017M610694. The special GPU accelerated supercomputer laohu at the Center of Information and Computing at National Astronomical Observatories, Chinese Academy of Sciences, funded by Ministry of Finance of Peoples Republic of China under grant ZDYZ2008-2 has been used for the computer simulations.

\def\aj{AJ} \def\actaa{Acta Astron.}  \def\araa{ARA\&A} \def\apj{ApJ}
\def\apjl{ApJ} \def\apjs{ApJS} \def\ao{Appl.~Opt.}  \def\apss{Ap\&SS}
\def\aap{A\&A} \def\aapr{A\&A~Rev.}  \def\aaps{A\&AS} \def\azh{AZh}
\def\baas{BAAS} \def\bac{Bull. astr. Inst. Czechosl.}
\def\caa{Chinese Astron. Astrophys.}  \def\cjaa{Chinese
  J. Astron. Astrophys.}  \def\icarus{Icarus} \def\jcap{J. Cosmology
  Astropart. Phys.}  \def\jrasc{JRASC} \def\mnras{MNRAS}
\def\memras{MmRAS} \def\na{New A} \def\nar{New A Rev.}
\def\pasa{PASA} \def\pra{Phys.~Rev.~A} \def\prb{Phys.~Rev.~B}
\def\prc{Phys.~Rev.~C} \def\prd{Phys.~Rev.~D} \def\pre{Phys.~Rev.~E}
\def\prl{Phys.~Rev.~Lett.}  \def\pasp{PASP} \def\pasj{PASJ}
\def\qjras{QJRAS} \def\rmxaa{Rev. Mexicana Astron. Astrofis.}
\def\skytel{S\&T} \def\solphys{Sol.~Phys.}  \def\sovast{Soviet~Ast.}
\def\ssr{Space~Sci.~Rev.}  \def\zap{ZAp} \def\nat{Nature}
\def\natas{Nature Astronomy}
\def\iaucirc{IAU~Circ.}  \def\aplett{Astrophys.~Lett.}
\def\apspr{Astrophys.~Space~Phys.~Res.}
\def\bain{Bull.~Astron.~Inst.~Netherlands}
\def\fcp{Fund.~Cosmic~Phys.}  \def\gca{Geochim.~Cosmochim.~Acta}
\def\grl{Geophys.~Res.~Lett.}  \def\jcp{J.~Chem.~Phys.}
\def\jgr{J.~Geophys.~Res.}
\def\jqsrt{J.~Quant.~Spec.~Radiat.~Transf.}
\def\memsai{Mem.~Soc.~Astron.~Italiana} \def\nphysa{Nucl.~Phys.~A}
\def\physrep{Phys.~Rep.}  \def\physscr{Phys.~Scr}
\def\planss{Planet.~Space~Sci.}  \def\procspie{Proc.~SPIE}
\let\astap=\aap \let\apjlett=\apjl \let\apjsupp=\apjs \let\applopt=\ao

\bibliographystyle{mn} 
\bibliography{mnras17b}
\label{lastpage}

\end{document}